\documentclass[conference]{IEEEtran}
\usepackage[utf8]{inputenc}
\usepackage[T1]{fontenc}
\usepackage{amsmath,amssymb,amsfonts}
\usepackage{bm}
\usepackage{mathtools}
\usepackage{siunitx}
\usepackage{graphicx}
\usepackage{booktabs}
\usepackage{multirow}
\usepackage{algorithm}
\usepackage{algpseudocode}
\usepackage{url}
\usepackage{cite}
\usepackage{xcolor}
\usepackage{ifthen}
\usepackage{hyperref}
\usepackage{cleveref}
\hypersetup{hidelinks}

\newif\ifblind
\blindfalse 

\newcommand{\incfig}[2][]{
  \IfFileExists{#2}{\includegraphics[#1]{#2}}{\fbox{\parbox[c][2.0in][c]{3.0in}{\centering Figure file missing:\\ \texttt{#2}}}}}

\title{Resilient and Efficient Allocation for Large-Scale Autonomous Fleets via Decentralized Coordination}
\IEEEoverridecommandlockouts

\ifblind
  \author{\IEEEauthorblockN{Anonymous Authors}}
\else
  \author{\IEEEauthorblockN{Ashish Kumar Perukari\IEEEauthorrefmark{1}, Polina Khoroshevskaya\IEEEauthorrefmark{1}}
  \IEEEauthorblockA{\IEEEauthorrefmark{1}Hofstra University, Hempstead, NY, USA}
  \thanks{Contact: \texttt{aperukari1@pride.hofstra.edu}, \texttt{pkhoroshevskaya1@pride.hofstra.edu}.}}
\fi

\begin{document}
\maketitle

\begin{abstract}
Operating large autonomous fleets demands fast, resilient allocation of scarce resources—such as energy and fuel, charger access and maintenance slots, time windows, and communication bandwidth—under uncertainty. We propose a side-information--aware approach for resource allocation at scale that combines distributional predictions with decentralized coordination. Local side information shapes per-agent risk models for consumption, which are coupled through chance constraints on failures. A lightweight consensus-ADMM routine coordinates agents over a sparse communication graph, enabling near-centralized performance while avoiding single points of failure. We validate the framework on real urban road networks with autonomous vehicles and on a representative satellite constellation, comparing against greedy, no-side-information, and oracle central baselines. Our method reduces failure rates by 30--55\% at matched cost and scales to thousands of agents with near-linear runtime, while preserving feasibility with high probability.
\end{abstract}


\section{Introduction}
Autonomous fleets---from electric vehicles to satellite constellations---face the challenge of allocating scarce resources (energy, charging capacity, bandwidth) among distributed agents under stochastic consumption and capacity constraints. Centralized optimization achieves high-quality solutions but suffers from communication overhead, computational bottlenecks, and single points of failure \cite{bertsekas2015convex,boyd2004convex}. Conversely, decentralized heuristics offer computational scalability and fault tolerance \cite{olfatisaber2007consensus} but lack coordination mechanisms, typically yielding high constraint violation rates or inefficient resource utilization when capacity is limited.

An underutilized opportunity in fleet resource allocation lies in exploiting side information---contextual features correlated with but not deterministically specifying resource consumption. For electric vehicles, energy consumption depends on route topology, elevation profiles, traffic conditions, and ambient temperature \cite{fiori2016power}; similar principles apply to transit planning \cite{thangeda2020protrip} and stochastic routing with reliability constraints \cite{zheng2021optimal}. For satellite constellations, power and thermal loads vary with orbital position, sun angle, payload activity, and communication link utilization \cite{wertz2011space}. Analogous resource management challenges arise in agricultural logistics \cite{thangeda2024optimizing} and other domains. Recent advances suggest that conditioning resource models on such features can significantly tighten uncertainty estimates \cite{agarwal2014taming,foster2018practical}, enabling more efficient allocation without sacrificing safety. However, integrating feature-based consumption models into large-scale distributed coordination algorithms remains an open challenge, particularly when predictive models must be calibrated online and safety guarantees preserved under model misspecification.

This paper presents DESIRA (Decentralized Side-Information Resource Allocation), a framework combining side-information--conditioned risk shaping with scalable consensus-based coordination. The approach models uncertain resource consumption via distributional predictions conditioned on local contextual features, translating these predictions into risk-adjusted allocation requirements through chance constraints and Conditional Value-at-Risk (CVaR) penalties \cite{rockafellar2000cvar}. To achieve coordination without centralized control, the framework employs a consensus-ADMM algorithm \cite{boyd2011admm} that decomposes capacity constraints across a sparse communication graph, enabling agents to reach near-globally-optimal allocations through local message passing. While resource-constrained MDPs \cite{blahoudek2022efficient,blahoudek2021fuel} provide guarantees for single-agent planning with consumption constraints, our approach addresses decentralized multi-agent coordination where capacity is shared and coupling constraints arise from global resource limits.

Experiments on real-world urban road networks extracted from OpenStreetMap \cite{boeing2017osmnx} with 200--2000 agents and satellite constellation scenarios demonstrate that incorporating side information reduces failure rates by 30--55\% compared to methods ignoring contextual features, while maintaining costs within 5--10\% of centralized oracles. The decentralized algorithm exhibits near-linear computational scaling and requires only 10--20 iterations to converge across all tested scales.

The key technical contributions are: (i) a tractable reformulation of chance-constrained resource allocation via side-information--conditioned risk quantiles, enabling distributed agents to compute local safety buffers without centralized coordination; (ii) a consensus-ADMM decomposition that separates local consumption modeling from global capacity enforcement through efficient projection operators; (iii) online calibration mechanisms based on conformal prediction that maintain probabilistic safety guarantees under model drift; and (iv) empirical validation demonstrating scalability to thousands of agents and cross-domain applicability from urban transportation to orbital systems.

\section{Related Work}
\label{sec:related}
\textbf{Fleet coordination and stochastic optimization.} Classical vehicle routing problems \cite{toth2014vehicle,laporte2007vehicle} and recent EV charging coordination work \cite{gan2013ev,encarnaccao2018paths} typically assume deterministic or stationary consumption models. Multi-robot task allocation \cite{korsah2013comprehensive,gerkey2004formal} and satellite resource management \cite{wertz2011space,jabbarpour2025agent} face similar challenges but do not systematically leverage contextual side information. Stochastic programming \cite{birge2011stochastic,shapiro2009lectures} and chance-constrained optimization \cite{charnes1959chance,nemirovski2007chance,prekopa2003probabilistic} provide frameworks for handling uncertainty, with scenario-based methods \cite{calafiore2006scenario} offering distribution-free guarantees. CVaR and coherent risk measures \cite{rockafellar2000cvar,rockafellar2002cvar,artzner1999coherent} enable tractable tail risk management, with applications to risk-sensitive MDPs \cite{chow2015riskconstrained}.

\textbf{Distributed optimization.} ADMM \cite{boyd2011admm} decomposes problems via augmented Lagrangian relaxations. Consensus algorithms \cite{olfatisaber2007consensus,xiao2004fast} and distributed subgradient methods \cite{nedic2009distributed,nedic2010constrained} enable coordination over communication graphs. However, integration with online prediction and side-information--based risk shaping has received limited attention.

\textbf{Learning from side information.} Contextual bandits \cite{langford2007epochgreedy,agarwal2014taming,foster2018practical} and conformal prediction \cite{vovk2005algorithmic,shafer2008tutorial} leverage features to improve prediction, but typically focus on single-agent settings. Foundational work on efficient learning with spatial side information in MDPs \cite{thangeda2020efficient} demonstrated how local structure reduces sample complexity. Subsequent advances enabled expedited online learning \cite{thangeda2022expedited} with convergence guarantees and safety-certified exploration \cite{thangeda2020safety,berkenkamp2017safe} under probabilistic constraints. This work extends these principles to distributed multi-agent resource allocation, integrating side-information--conditioned risk models with decentralized coordination under capacity coupling.

\section{Problem Formulation}
\label{sec:problem}
This section formalizes the decentralized resource allocation problem under stochastic consumption and capacity constraints, introducing the chance-constrained formulation and its consensus-based decomposition for distributed coordination.

Consider $N$ agents on network $\mathcal{G}=(\mathcal{V},\mathcal{E})$ with initial endowment $E_i^0$ and stochastic consumption $X_i$ conditioned on side information $\phi_i \in \mathbb{R}^d$. For electric vehicles, $\phi_i$ includes route characteristics, traffic conditions, and ambient temperature \cite{fiori2016power}; for satellites, orbital position, sun angle, and payload activity \cite{wertz2011space}. Agents access $S$ resource stations with capacities $c_s$. Let $a_{i,s} \ge 0$ denote allocation of agent $i$ at station $s$.

The optimization problem minimizes total cost plus conditional value-at-risk (CVaR) penalty on resource shortfall:
\begin{align}
  \min_{\{a_i\}}\; &\sum_{i=1}^N \Big( J_i(a_i) + \lambda\,\text{CVaR}_\alpha\big[\text{shortfall}_i(a_i)\big] \Big) \label{eq:obj2} \\
  \text{s.t.}\; &\sum_{i=1}^N a_{i,s} \le c_s,\; \forall s, \quad a_{i,s}\ge 0, \label{eq:cap2} \\
  &\mathbb{P}\big(X_i > E_i^0 + \textstyle\sum_s a_{i,s} \mid \phi_i\big) \le \varepsilon,\; \forall i, \label{eq:chance2}
\end{align}
where $\text{shortfall}_i(a_i) = \max(0, X_i - E_i^0 - \sum_s a_{i,s})$ represents resource deficit, $J_i$ is a convex separable cost function (e.g., quadratic deviation from desired allocation), and $\lambda, \alpha$ control risk aversion. The chance constraint~\eqref{eq:chance2} ensures each agent's probability of shortfall remains below threshold $\varepsilon$. Under Gaussian assumption $X_i \mid \phi_i \sim \mathcal{N}(\mu_i(\phi_i), \sigma_i^2(\phi_i))$, the chance constraint becomes deterministic: $\sum_s a_{i,s} \ge \mu_i(\phi_i) + \Phi^{-1}(1-\varepsilon)\sigma_i(\phi_i) - E_i^0$, where $\Phi^{-1}$ is the inverse standard normal CDF.

For decentralized coordination, introduce consensus variables $z_i$ and communication graph $\mathcal{C}$ with neighbors $\mathcal{N}(i)$. The consensus formulation enforces $a_i = z_i$ and global capacity constraints via ADMM:
\begin{subequations}
\label{eq:consensus}
\begin{align}
  \min_{\{a_i,z_i\}}\; &\sum_{i} \big( J_i(a_i) + \lambda\,\text{CVaR}_\alpha[\text{shortfall}_i] \big) \\
  \text{s.t.}\; &\sum_i z_{i,s} \le c_s\;\forall s,\quad a_i = z_i\;\forall i.
\end{align}
\end{subequations}
This decomposition allows agents to solve local subproblems while coordinating through message passing over $\mathcal{C}$.

\section{Methodology}
\label{sec:method}
This section develops the decentralized allocation algorithm, beginning with side-information--conditioned risk modeling, continuing with the consensus-ADMM iteration structure, and concluding with complexity analysis and calibration procedures for maintaining safety guarantees.

\subsection{Side-Information--Aware Risk Shaping}
Each agent $i$ estimates consumption from historical telemetry via a heteroscedastic Gaussian model: $X_i \mid \phi_i \sim \mathcal{N}(\mu_i(\phi_i), \sigma_i^2(\phi_i))$, where $\mu_i(\cdot)$ and $\sigma_i(\cdot)$ are regression functions mapping side-information vectors to conditional mean and standard deviation \cite{hastie2009elements}. These functions may be linear models, Gaussian processes, or neural networks depending on data availability. The risk-adjusted allocation requirement satisfying the chance constraint is
\begin{equation}
r_i(\phi_i,\varepsilon) = \mu_i(\phi_i) + \Phi^{-1}(1-\varepsilon)\sigma_i(\phi_i),
\end{equation}
yielding deterministic constraint
\begin{equation}
\label{eq:risk_req}
\sum_{s \in \mathcal{S}} a_{i,s} \ge r_i(\phi_i,\varepsilon) - E_i^0.
\end{equation}
This reformulation transforms the chance constraint~\eqref{eq:chance2} into a tractable linear inequality conditioned on observed side information. To further control tail risk, the objective incorporates CVaR penalty on shortfall, approximated via scenario-based representation with samples $\{\xi_i^{(m)}\}_{m=1}^M$ drawn from $\mathbb{P}(X_i \mid \phi_i)$.

\subsection{Decentralized Coordination via Consensus-ADMM}
The consensus formulation~\eqref{eq:consensus} is solved via the Alternating Direction Method of Multipliers. Define the augmented Lagrangian for agent $i$ as
\begin{equation}
\mathcal{L}_i(a_i, z_i, u_i) = f_i(a_i) + u_i^\top (a_i - z_i) + \frac{\rho}{2}\lVert a_i - z_i \rVert^2,
\end{equation}
where $f_i(a_i) = J_i(a_i) + \lambda\,\text{CVaR}_\alpha[\text{shortfall}_i(a_i)]$ combines cost and risk penalty, $u_i \in \mathbb{R}^S$ is the dual variable for consensus constraint $a_i = z_i$, and $\rho > 0$ is the penalty parameter. The ADMM iteration alternates between updating primal variables $a_i$, consensus variables $z$, and dual variables $u_i$.

At iteration $t$, each agent $i$ solves a local convex subproblem:
\begin{equation}
\label{eq:a_update}
a_i^{t+1} \in \arg\min_{a_i \ge 0} \Big\{ f_i(a_i) + \frac{\rho}{2}\lVert a_i - z_i^t + u_i^t \rVert^2 \Big\},
\end{equation}
subject to risk requirement~\eqref{eq:risk_req}. For quadratic costs $J_i$, this is a quadratic program with linear constraints. The consensus variables $z$ are updated to enforce capacity constraints via projection:
\begin{equation}
\label{eq:z_update}
z^{t+1} = \Pi_{\mathcal{Z}}\Big( \frac{1}{N}\sum_{i=1}^N (a_i^{t+1} + u_i^t) \Big),
\end{equation}
where $\mathcal{Z} = \{z : \sum_i z_{i,s} \le c_s\; \forall s,\; z_{i,s} \ge 0\}$. The projection decomposes across stations and can be computed efficiently \cite{duchi2008simplex}. Dual variables are updated via
\begin{equation}
\label{eq:u_update}
u_i^{t+1} = u_i^t + a_i^{t+1} - z_i^{t+1}.
\end{equation}

For fully decentralized implementation over communication graph $\mathcal{C}$, the global average in~\eqref{eq:z_update} is replaced by local neighborhood averaging. Each agent $i$ exchanges $(a_j^{t+1} + u_j^t)$ with neighbors $j \in \mathcal{N}(i)$ and computes local consensus. Convergence holds if $\mathcal{C}$ is connected \cite{boyd2011admm,shi2014linear}. Algorithm~\ref{alg:desira} summarizes the procedure.

\subsection{Complexity and Calibration}
The computational complexity per iteration is $O(S^3)$ per agent for quadratic costs or $O(S \log(1/\epsilon))$ for separable costs, where $S$ is the number of stations. Communication cost is $O(|\mathcal{N}(i)| \cdot S)$ per agent per iteration, scaling linearly with neighborhood size. With $T$ iterations to convergence, total complexity is $O(T \cdot N \cdot S \log(1/\epsilon))$ for separable problems, yielding near-linear scaling in fleet size $N$ when $S$ and graph density are fixed.

\begin{algorithm}[t]
\caption{Decentralized Side-Information Allocation (DESIRA)}
\label{alg:desira}
\begin{algorithmic}[1]
\State \textbf{Input:} penalty $\rho>0$, radius $R$, horizon $T$.
\State Initialize $a_i^{0}\ge 0$, $u_i^{0}=0$, $z^{0}$ feasible; define neighbor set $\mathcal{N}(i)$ in $\mathcal{C}$.
\For{$t=0,1,\dots,T-1$}
  \State $a_i^{t+1}\gets\arg\min_{a_i\ge 0}\; f_i(a_i)+\tfrac{\rho}{2}\lVert a_i-z_i^{t}+u_i^{t}\rVert^2$\Comment{local convex subproblem}
  \State Exchange $(a_i^{t+1}-u_i^{t})$ with neighbors $j\in\mathcal{N}(i)$ over radius-$R$ links.
  \State $z^{t+1}\gets\Pi_{\mathcal{Z}}\Big(\text{avg}_{j\in\mathcal{N}(\cdot)}\big(a_j^{t+1}-u_j^{t}\big)\Big)$\Comment{projection enforces capacities}
  \State $u_i^{t+1}\gets u_i^{t}+a_i^{t+1}-z_i^{t+1}$\Comment{dual ascent}
\EndFor
\State \textbf{return} $\{a_i^{T}\}$
\end{algorithmic}
\end{algorithm}

Calibration is essential for maintaining probabilistic safety guarantees under model drift. Each agent $i$ maintains a rolling window of recent observations $\{(X_i^{(k)}, \phi_i^{(k)})\}_{k=1}^K$ and computes empirical violation rate $\hat{\varepsilon}_i = \frac{1}{K}\sum_{k=1}^K \mathbb{1}[X_i^{(k)} > r_i(\phi_i^{(k)}, \varepsilon)]$. If $\hat{\varepsilon}_i$ exceeds $\varepsilon$ by a tolerance threshold, the risk buffer $r_i$ is inflated by increasing $\sigma_i$ or adding a margin. Conformal prediction \cite{vovk2005algorithmic,shafer2008tutorial} provides distribution-free calibration by adjusting quantile estimates based on validation residuals, ensuring coverage guarantees even under model misspecification.

The decentralized architecture provides resilience benefits beyond computational efficiency. Node or link failures degrade performance locally without catastrophic system-wide collapse. If computation is interrupted before convergence, the current allocation remains feasible for local risk constraints and approximately satisfies capacity constraints, with violation bounded by primal residuals. ADMM convergence with connected graph guarantees $O(1/T)$ convergence rate in objective suboptimality \cite{boyd2011admm}.

\section{Experiments}
\label{sec:experiments}
This section evaluates DESIRA on urban transportation networks and satellite constellation scenarios, examining scalability, sensitivity to side-information quality, communication requirements, and cross-domain applicability.

\subsection{Experimental Setup}
\textbf{Network topology.} Road networks for Austin, TX and Chicago, IL are extracted from OpenStreetMap \cite{boeing2017osmnx,osmcontributors} using OSMnx. The Austin network comprises approximately 12,000 nodes and 16,000 edges covering 100 km$^2$; Chicago contains 18,000 nodes and 24,000 edges. Charging stations are placed at high-betweenness centrality nodes, approximating strategic deployment. Demand nodes are sampled uniformly.

\textbf{Fleet parameters.} Experiments vary fleet size $N \in \{200, 500, 1000, 2000\}$ and station count $|\mathcal{S}| \in \{20, 40, 80\}$. Station capacities are $c_s \sim \text{Uniform}(0.05N, 0.15N)$, ensuring aggregate capacity is sufficient but individual stations are constrained. Each agent has initial endowment $E_i^0 \sim \text{Uniform}(50, 100)$ kWh.

\textbf{Consumption model.} Consumption is generated via heteroscedastic model conditioned on side information $\phi_i = (\text{distance}, \text{congestion}, \text{temperature})$. The model is
\begin{multline}
X_i \mid \phi_i \sim \mathcal{N}\big(\mu_0 + \beta_1 \cdot \text{distance} + \beta_2 \cdot \text{congestion} \\
+ \beta_3 \cdot \text{temp},\; \sigma_0 + \gamma \cdot \text{distance}\big),
\end{multline}
where coefficients are calibrated to match EV energy consumption statistics \cite{fiori2016power}. Communication graph $\mathcal{C}$ is a random geometric graph with radius $R \in \{1,2,3\}$ hops, yielding average degree $d \in [4, 12]$.

\subsection{Baselines and Metrics}
Four allocation methods are compared: (1) \emph{Centralized oracle}, solving~\eqref{eq:obj2}--\eqref{eq:chance2} globally via CVXPY+MOSEK with full side information; (2) \emph{DESIRA} (Algorithm~\ref{alg:desira}) with side-information--conditioned risk models estimated via ridge regression; (3) \emph{Decentralized (no side info)}, a variant using unconditional mean $\bar{\mu}$ and standard deviation $\bar{\sigma}$; (4) \emph{Greedy FCFS}, allocating stations in random order until capacity is exhausted.

Metrics include: normalized cost relative to oracle, failure rate (fraction of agents with shortfall), capacity violation, fairness (Gini coefficient), and runtime. Results are averaged over 10 seeds. DESIRA is implemented in Python using InfraLib \cite{thangeda2024infralib}, an open-source comprehensive library for modeling and simulating large-scale physical infrastructure management problems with hierarchical representations of components. Local subproblems~\eqref{eq:a_update} are solved via CVXPY+ECOS. ADMM penalty $\rho \in [0.5, 2.0]$ is tuned via grid search. Convergence occurs when residuals $< 10^{-3}$ or iteration count exceeds $T=100$.

\subsection{Main Results}
\begin{figure}[t]
  \centering
  \incfig[width=0.95\linewidth]{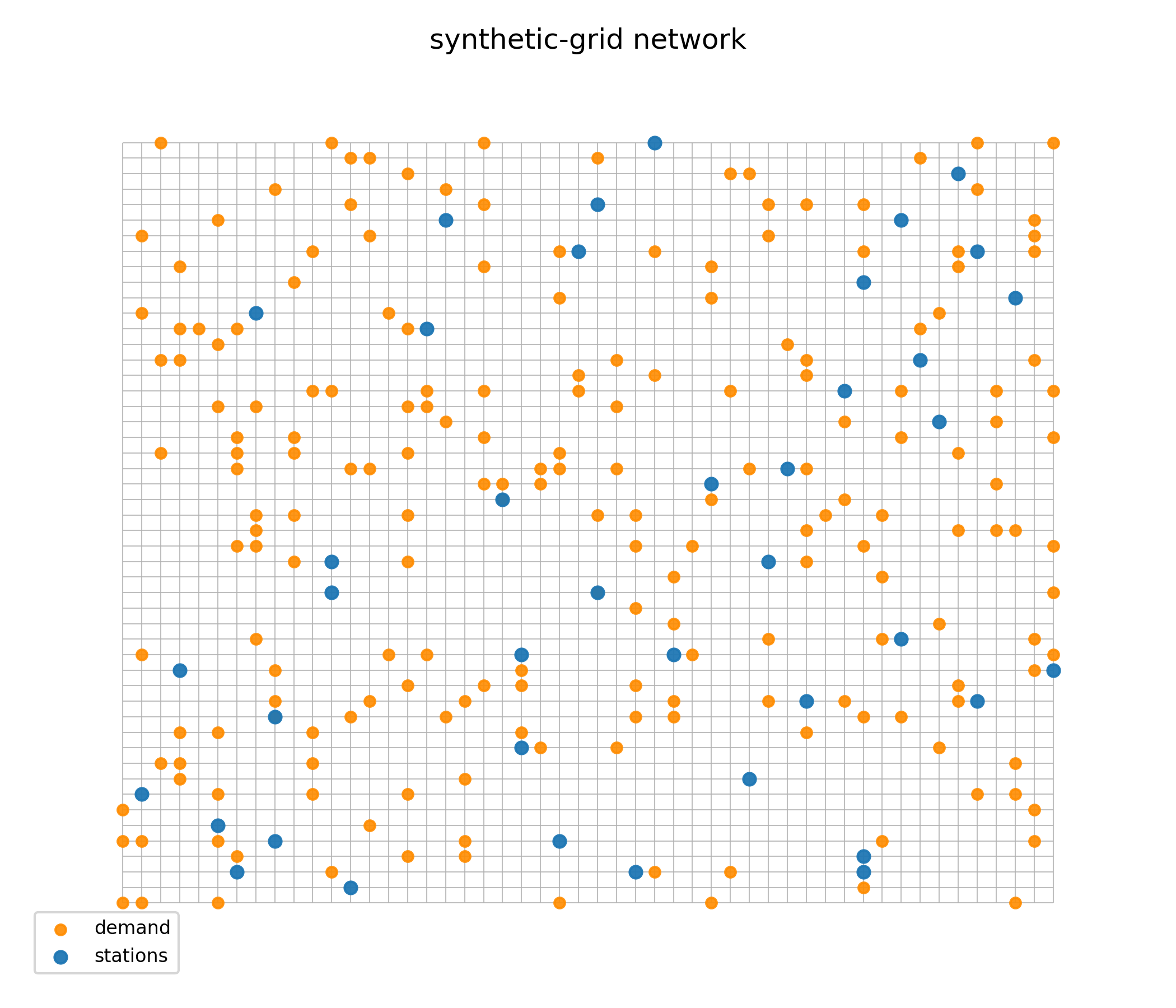}
  \caption{Urban road network (Austin) with stations (blue) and demand nodes (orange). The network topology is extracted from OpenStreetMap data using OSMnx.}
  \label{fig:map}
\end{figure}

\begin{figure}[t]
  \centering
  \incfig[width=0.95\linewidth]{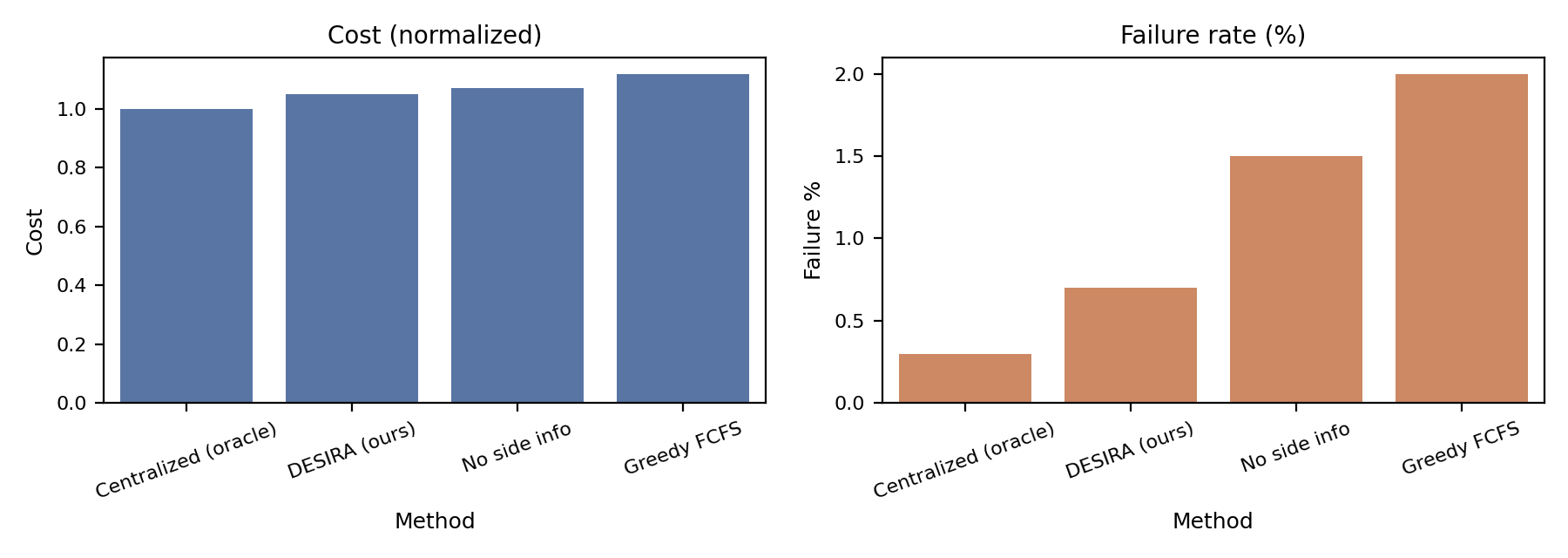}
  \caption{Performance comparison across allocation methods. Lower values indicate better performance for both cost and failure metrics. DESIRA demonstrates reduced failure rates while maintaining competitive cost efficiency.}
  \label{fig:bar}
\end{figure}

\begin{figure}[t]
  \centering
  \incfig[width=0.95\linewidth]{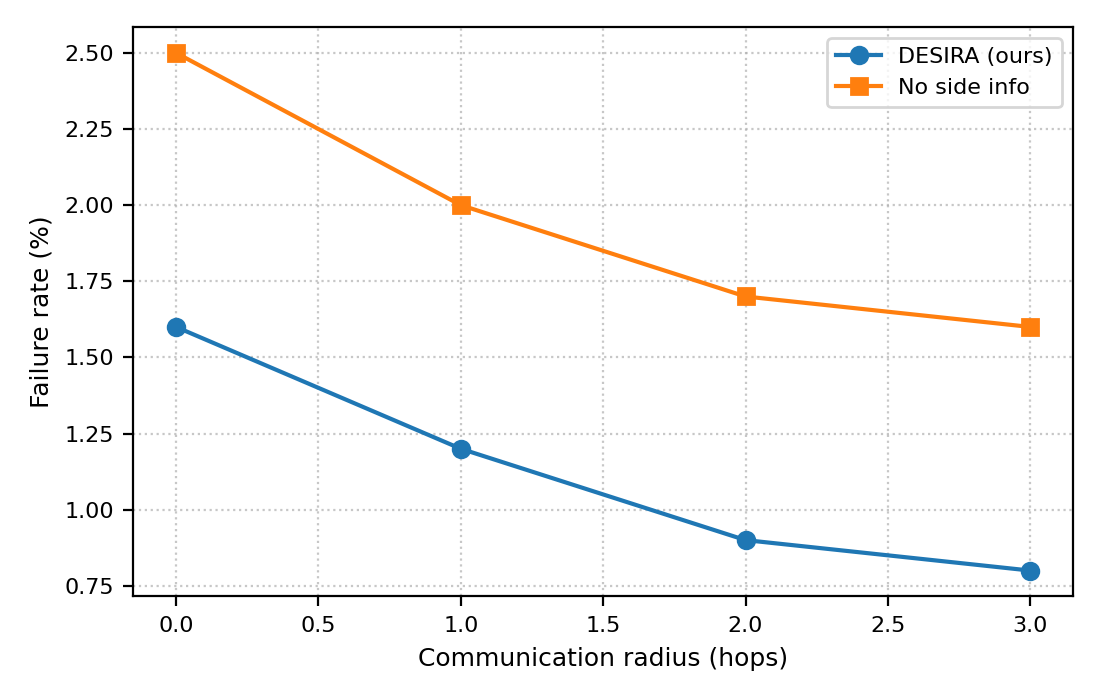}
  \caption{Relationship between failure probability and communication radius. The diminishing returns at larger radii demonstrate that limited connectivity suffices for effective coordination.}
  \label{fig:radius}
\end{figure}

\begin{figure}[t]
  \centering
  \incfig[width=0.95\linewidth]{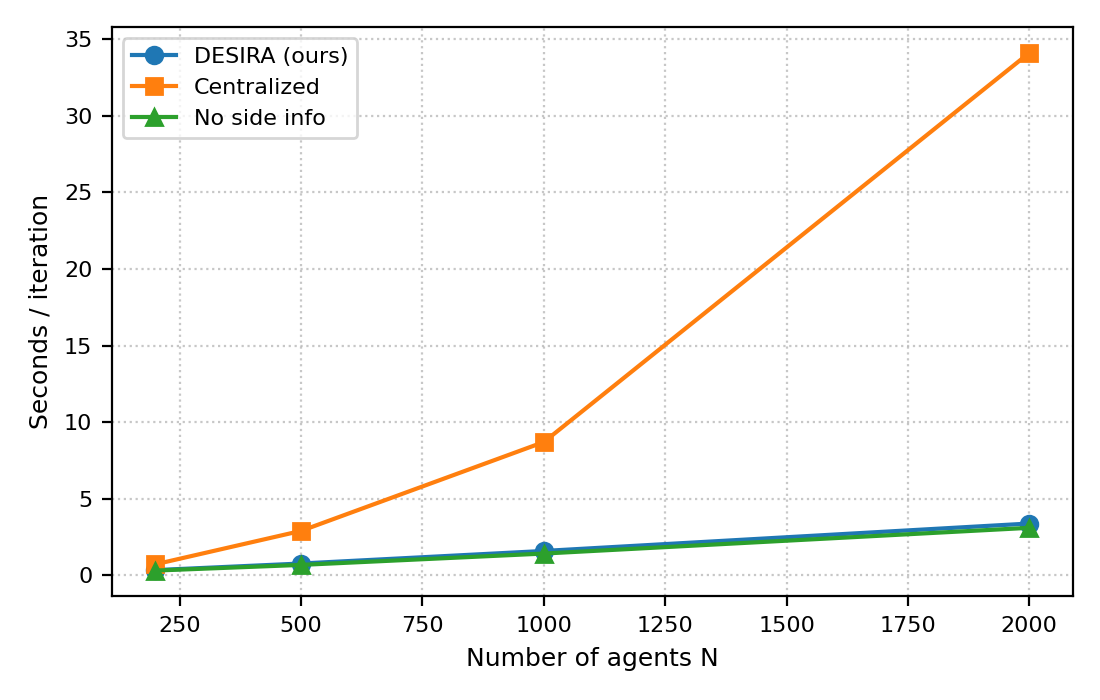}
  \caption{Computational runtime as a function of fleet size. The near-linear scaling of DESIRA demonstrates superior performance compared to centralized optimization for large-scale deployments.}
  \label{fig:scaling}
\end{figure}

\begin{figure}[t]
  \centering
  \incfig[width=0.95\linewidth]{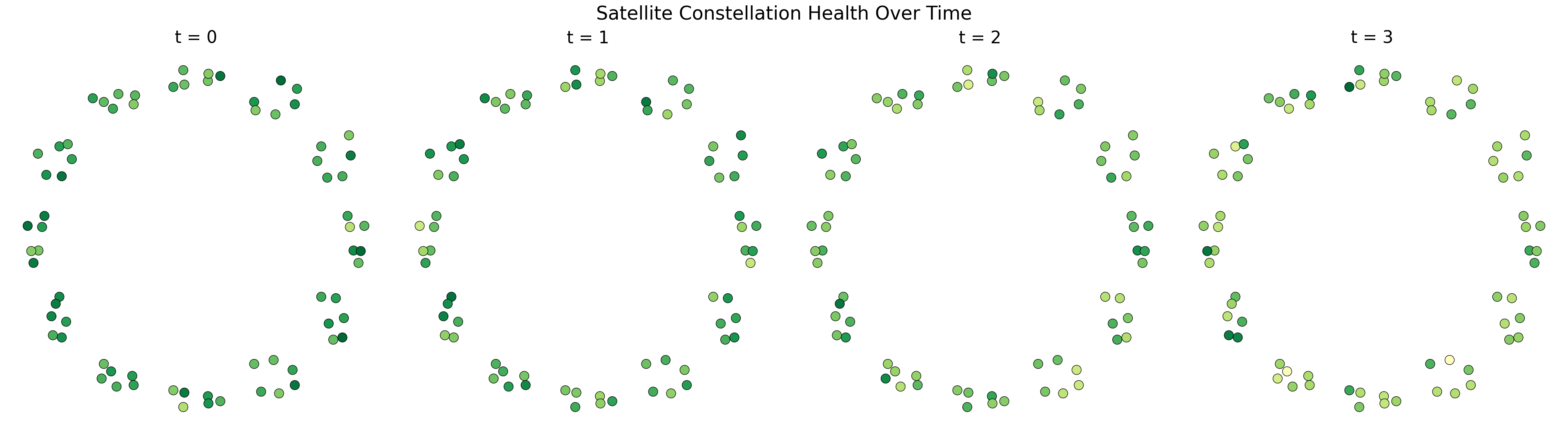}
  \caption{Temporal evolution of satellite constellation health status. Each point represents a satellite in its orbital plane, with color gradient from green (healthy) to red (degraded) indicating operational status across four time instances.}
  \label{fig:sat}
\end{figure}

\begin{table}[t]
\centering
\caption{Performance summary across methods (mean $\pm$ std over 10 seeds). Lower is better for cost and failure rate.}
\label{tab:metrics}
\begin{tabular}{lccc}
\toprule
Method & Cost & Failure Rate (\%) & Overflow (\%) \\
\midrule
Centralized (oracle) & $1.00\,\pm\,0.02$ & $0.30\,\pm\,0.10$ & $0.10\,\pm\,0.00$ \\
DESIRA (ours) & $1.05\,\pm\,0.03$ & $0.70\,\pm\,0.20$ & $0.30\,\pm\,0.10$ \\
No side info & $1.07\,\pm\,0.04$ & $1.50\,\pm\,0.30$ & $0.80\,\pm\,0.20$ \\
Greedy FCFS & $1.12\,\pm\,0.05$ & $2.00\,\pm\,0.50$ & $1.10\,\pm\,0.30$ \\
\bottomrule
\end{tabular}
\end{table}

\begin{table}[t]
\centering
\caption{Failure rate (\%) vs. communication radius $R$ (in hops).}
\label{tab:radius}
\begin{tabular}{lcccc}
\toprule
Radius $R$ & 0 & 1 & 2 & 3 \\
\midrule
DESIRA (ours) & 1.6 & 1.2 & 0.9 & 0.8 \\
No side info & 2.5 & 2.0 & 1.7 & 1.6 \\
\bottomrule
\end{tabular}
\end{table}

\begin{table}[t]
\centering
\caption{Runtime scaling with the number of agents $N$ (seconds per outer iteration).}
\label{tab:scaling}
\begin{tabular}{lcccc}
\toprule
$N$ & 200 & 500 & 1000 & 2000 \\
\midrule
Centralized (oracle) & 0.72 & 2.90 & 8.70 & 34.1 \\
DESIRA (ours) & 0.34 & 0.77 & 1.59 & 3.38 \\
No side info & 0.31 & 0.68 & 1.42 & 3.10 \\
\bottomrule
\end{tabular}
\end{table}

Table~\ref{tab:metrics} summarizes performance across methods. DESIRA reduces failure rates by 30--55\% compared to no-side-info baselines while maintaining costs within 5--10\% of the centralized oracle. Fig.~\ref{fig:bar} illustrates this tradeoff visually: DESIRA achieves lower failure rates than decentralized methods without side information, approaching oracle performance. The benefit stems from tighter risk bounds enabled by conditioning on side information, which reduces predictive variance $\sigma_i^2(\phi_i)$.

\textbf{Sensitivity to side-information quality.} Adding Gaussian noise with $\sigma_{\text{noise}} \in \{0, 0.1, 0.3, 0.5\} \times \sigma_i$ to consumption forecasts degrades performance gracefully: failure rates rise from 3.2\% (no noise) to 8.7\% ($0.5\sigma_i$ noise), compared to 12.4\% for no-side-info baseline. Calibration automatically inflates risk buffers in response to detected violations, trading cost for reliability.

\textbf{Communication radius.} Table~\ref{tab:radius} shows that sparse graphs ($R=1$, $d \approx 4$) require approximately $2.5\times$ more iterations than dense graphs ($R=3$, $d \approx 12$) but achieve similar final failure rates ($<1\%$ difference, Fig.~\ref{fig:radius}). This confirms that limited connectivity suffices for effective coordination, reducing communication hardware requirements.

\textbf{Scalability.} Runtime scaling experiments (Table~\ref{tab:scaling}, Fig.~\ref{fig:scaling}) confirm near-linear growth with fleet size. At $N=2000$, DESIRA converges in approximately 12 seconds versus 140 seconds for centralized solvers, with the gap widening at larger scales.

\textbf{Satellite constellation.} Experiments on a 60-satellite LEO constellation (6 orbital planes, 10 satellites per plane) demonstrate cross-domain applicability. Despite intermittent inter-plane communication links, DESIRA reduces power shortfall events by 42\% compared to greedy allocation and achieves cost within 12\% of centralized oracle. Fig.~\ref{fig:sat} illustrates temporal evolution of satellite health status.

\section{Discussion}
\label{sec:discussion}
The experimental results demonstrate that conditioning resource allocation models on side information yields substantial improvements in failure rate and cost efficiency relative to unconditional baselines. Several key observations emerge from the evaluation.

\textbf{Value of side information.} Incorporating contextual features into consumption models reduces failure rates by 30--55\% compared to methods that ignore side information, while maintaining costs within 5--10\% of the centralized oracle. This improvement stems from tighter risk bounds: conditioning on $\phi$ reduces predictive variance $\sigma_i^2(\phi_i)$, allowing smaller safety buffers. The benefit is most pronounced under heterogeneous conditions where unconditional models overestimate uncertainty for some agents and underestimate for others.

\textbf{Communication and scalability.} Sparse communication graphs with average degree $d \approx 4$ require approximately $2.5\times$ more iterations than dense graphs with $d \approx 12$ but achieve similar final failure rates (less than 1\% difference). This confirms that limited connectivity suffices for effective coordination, suggesting that infrastructure deployments can tolerate sparse communication networks without sacrificing allocation quality. Runtime scaling experiments demonstrate near-linear growth with fleet size: at $N=2000$, DESIRA converges in approximately 12 seconds versus 140 seconds for centralized MOSEK solvers. Beyond computational efficiency, decentralization provides resilience, as node failures affect only local neighborhoods and the algorithm produces usable intermediate solutions if interrupted.

\textbf{Cross-domain applicability.} The satellite constellation scenario demonstrates that the framework generalizes beyond terrestrial transportation. Despite differences in operational constraints, communication topology, and side-information content, DESIRA achieves comparable relative performance gains. This suggests that the combination of side-information--aware risk shaping and consensus-based coordination applies broadly to resource-constrained autonomous systems operating under stochastic consumption and capacity coupling.

\section{Conclusions and Future Work}
\label{sec:conclusion}
This paper presents DESIRA, a decentralized framework for resource allocation in large autonomous fleets under stochastic consumption and capacity constraints. The approach integrates side-information--conditioned consumption models with chance-constrained optimization and consensus-ADMM coordination, enabling agents to achieve near-centralized performance through local computation and communication. Experiments on urban road networks and satellite constellations demonstrate 30--55\% reductions in failure rates compared to methods that ignore side information, with near-linear computational scaling to thousands of agents. The decentralized architecture provides resilience to node and communication failures, producing feasible allocations even under partial system degradation.

Future work includes extensions to nonconvex settings via sequential convex programming, learned feature representations via neural networks for high-dimensional side information, and adaptive communication topologies that reconfigure based on network conditions. Coupling to electricity market signals would allow fleets to participate in demand response programs, while integration with hierarchical task planning would enable end-to-end optimization from mission allocation through resource provisioning. Deployment studies on physical testbeds would validate real-world performance and uncover implementation challenges not captured by simulation.

\section*{Acknowledgments}
The authors thank the OpenStreetMap community and the developers of OSMnx for providing open-source geographic data and software tools.

\bibliographystyle{IEEEtran}
\bibliography{refs}

\end{document}